\documentclass{SciPost}

\hypersetup{
    colorlinks,
    linkcolor={red!50!black},
    citecolor={blue!50!black},
    urlcolor={blue!80!black}
}

\usepackage[bitstream-charter]{mathdesign}
\DeclareSymbolFont{usualmathcal}{OMS}{cmsy}{m}{n}
\DeclareSymbolFontAlphabet{\mathcal}{usualmathcal}

\fancypagestyle{SPstyle}{
\fancyhf{}
\lhead{\colorbox{scipostblue}{\bf \color{white} ~SciPost Physics Core }}
\rhead{{\bf \color{scipostdeepblue} ~Submission }}

\fancyfoot[C]{\textbf{\thepage}}
}

\usepackage{cite}
\usepackage{graphicx}
\newcommand{\be}{\begin{equation}}
\newcommand{\ee}{\end{equation}}
\newcommand{\bea}{\begin{eqnarray}}
\newcommand{\eea}{\end{eqnarray}}

\usepackage{doi}

\begin{document}

\pagestyle{SPstyle}

\begin{center}{\Large \textbf{\color{scipostdeepblue}{
Effect of construction steels on PMTs detection efficiency at JUNO \\
}}}\end{center}

\begin{center}\textbf{
T.~Yan\textsuperscript{1},
J.~Songwadhana\textsuperscript{1},
A.~Limphirat\textsuperscript{1},
Y.~Yan\textsuperscript{1$\star$},
H.~Lu\textsuperscript{2},
F. Ning\textsuperscript{2},
P. Zheng\textsuperscript{2},
C.~Yang\textsuperscript{2},
G. Zhang\textsuperscript{3},
W.~Sreethawong\textsuperscript{1},
K.~Khosonthongkee\textsuperscript{1}, and
N.~Suwonjandee\textsuperscript{4} \\
on behalf of JUNO collaboration
}\end{center}

\begin{center}
{\bf 1} School of Physics and Center for Excellence in High Energy Physics and Astrophysics, Suranaree University of Technology,
111 University Avenue, Nakhon Ratchasima 30000, Thailand.
\\
{\bf 2} Institute of High Energy Physics, Chinese Academy of Sciences, 
Beijing 100049, China.
\\
{\bf 3} Institute of Advanced Science Facilities, Shenzhen, China.
\\
{\bf 4} Particle Physics Research Laboratory, Department of Physics, Faculty of Science, Chulalongkorn University, 254 Phayathai Rd., Patumwan, Bangkok 10330 Thailand.
\\[\baselineskip]
$\star$ \href{mailto:email1}{\small yupeng@sut.ac.th}\quad
\end{center}

\section*{\color{scipostdeepblue}{Abstract}}
\textbf{\boldmath{%
We study the impact of the carbon steel rebars and the steel TT bridge within the JUNO structure on the shielding effect of the coils. Our simulations demonstrate that despite the presence of carbon steel structures of the rebars of the water pool and the TT bridge within the central detector vicinity, the residual magnetic field experienced by the PMTs remains within the acceptable limit established by the JUNO experiment of 10\% for CD-PMTs and 20\% for Veto-PMTs, compared to the geomagnetic field.  The maximum magnetic fields experienced by the CD-PMTs and Veto-PMTs are 9\% and 18\% of the geomagnetic field strength, respectively. These findings indicate that the residual magnetic field has minimal impacts on the PMTs detection efficiency.
}}

\vspace{\baselineskip}

\noindent\textcolor{white!90!black}{%
\fbox{\parbox{0.975\linewidth}{%
\textcolor{white!40!black}{\begin{tabular}{lr}%
  \begin{minipage}{0.6\textwidth}%
    {\small Copyright attribution to authors. \newline
    This work is a submission to SciPost Physics Core. \newline
    License information to appear upon publication. \newline
    Publication information to appear upon publication.}
  \end{minipage} & \begin{minipage}{0.4\textwidth}
    {\small Received Date \newline Accepted Date \newline Published Date}%
  \end{minipage}
\end{tabular}}
}}
}


\vspace{10pt}
\noindent\rule{\textwidth}{1pt}
\tableofcontents
\noindent\rule{\textwidth}{1pt}
\vspace{10pt}


\section{Introduction}

\begin{figure}[h]
\centering 
\includegraphics[width=0.85\linewidth]{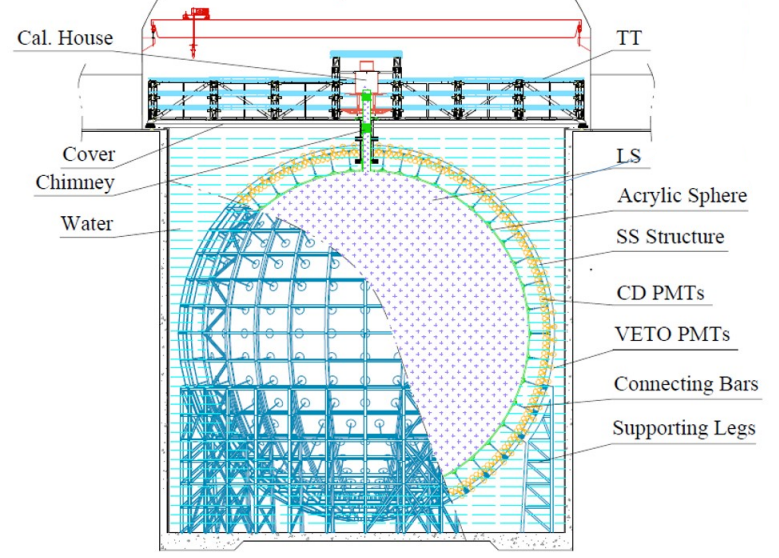}
\caption{\label{fig:layout} Layout of the JUNO experiment. The simulated ferromagnetic steel components are the top tracker (TT) bridge and rebars inside the concrete that forms the water pool. The supporting legs and the sphere structure are constructed from stainless steel, and thus not contribute magnetically. The components sensitive to magnetic field strength are the CD and Veto PMTs. Not shown in the drawing are the compensation coils wrapped around the central detector sphere \cite{Abusleme_2022}.}
\end{figure}

The JUNO detector comprises a central detector (CD), containing 17,612 20-inch Microchannel Plate Photomultiplier tubes (MCP-PMTs) and Hamamatsu PMTs~\cite{PhysRevD.88.013008}, and 25,600 3-inch PMTs outside an acrylic sphere of 35.4 m in diameter, which is filled with 20 kilotons of liquid scintillator, as shown in figure \ref{fig:layout}. Additionally, 2,400 20-inch MCP-PMT facing outwards to the pure outer water serve as a water Cherenkov detector or Veto-PMT. The 20-inch PMT of the central detector must achieve a scintillation light collection of at least 1,200
p.e./MeV and $3\%$ @ 1MeV energy resolution for physics requirement to observe inverse beta decay spectrum. Calibration simulations by the JUNO Collaboration showed a scintillation light collection of 1,655 p.e./MeV \cite{Abusleme_2025}. Achieving these requirements requires minimizing the geomagnetic field experienced by 20-inch PMTs, Hamamatsu PMT and MCP-PMT, to maintain homogeneous photoelectron collection efficiency~\cite{Abusleme_2021}. 
\begin{figure}[htbp]
\centering 
\includegraphics[width=0.85\linewidth]{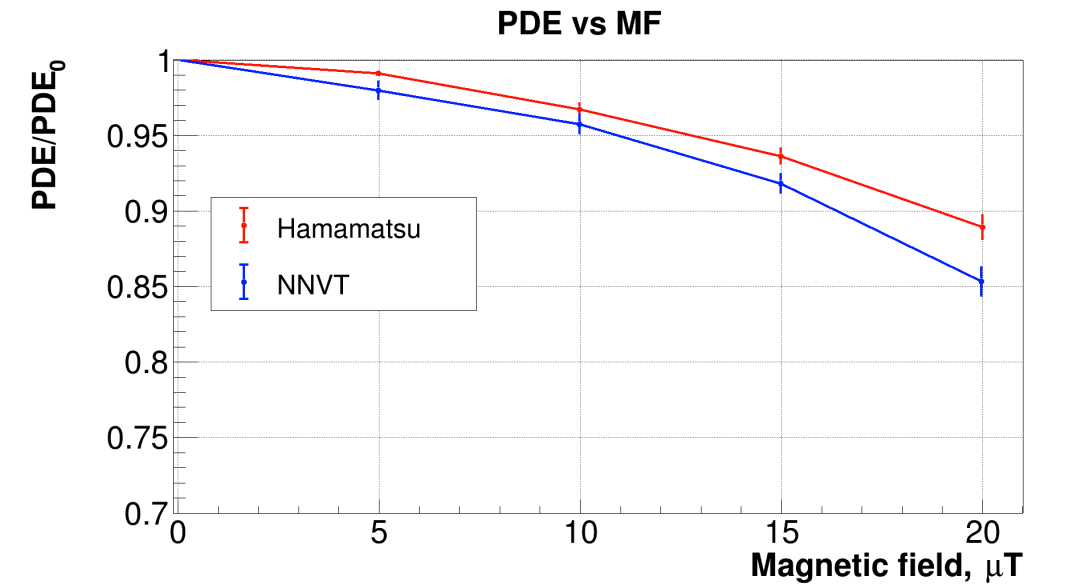}
\caption{\label{fig:pmteff} Averaged PDE versus magnetic field strength, tested with 9 HPK and 15 NNVT PMTs. The geomagnetic field strength at JUNO site is roughly 50$\mu T$ \cite{Abusleme_2022}.}
\end{figure}

The Lorentz force induced by the geomagnetic field can deflect photoelectron trajectories in these PMTs, resulting in a loss of photon detection efficiency (PDE), particularly evident in MCP-PMTs where the efficiency loss reaches approximately $15\%$ when the magnetic field intensity reaches 20 $\mu T$ (40\% of geomagnetic field strength), as illustrated in Figure~\ref{fig:pmteff}, which displays the PMT efficiency measured against the residual magnetic field intensity. The measured PDE of Hamamatsu PMT is represented by red points, while the measurement of MCP-PMT is represented by blue points. With a geomagnetic field strength below 5 $\mu T$, there is minimal loss of PDE. To achieve this, the JUNO experiment employs a set of 16 pairs of circular coils to compensate for the geomagnetic field within the detection region~\cite{WANG2012113,Zhang_2021}. These compensation coils are arranged according to the geometry of the central detector, with the objective of reducing the magnetic field to less than $10\%$ of the intensity of the geomagnetic field in the CD-PMT region between 38.5 and 39.5 m in diameter and below $20\%$ in the Veto-PMT area between 40.6 and 41.1 m in diameter.

\indent In Ref.~\cite{WANG2012113}, however, the effect of carbon steel which is an important component in the construction of the water pool and the top tracker (TT) bridge has not been considered. In this work, we study the impact of the carbon steel rebars and the steel TT bridge within the JUNO structure on the shielding effect of the coils. The paper is arranged as follows: In Section 2, we describe the layouts of the TT bridge as well as the rebars in the bottom and wall of the water pool. The TT bridge structure is rather complicated, one needs to simplify it for a calculation. We give an up-limit simplification in the section. The results are shown in Section 3, and the conclusions are given in Section 4.

\section{Description of Rebars and TT Bridge}
\begin{figure}[htbp]
\centering 
\includegraphics[width=0.6\linewidth]{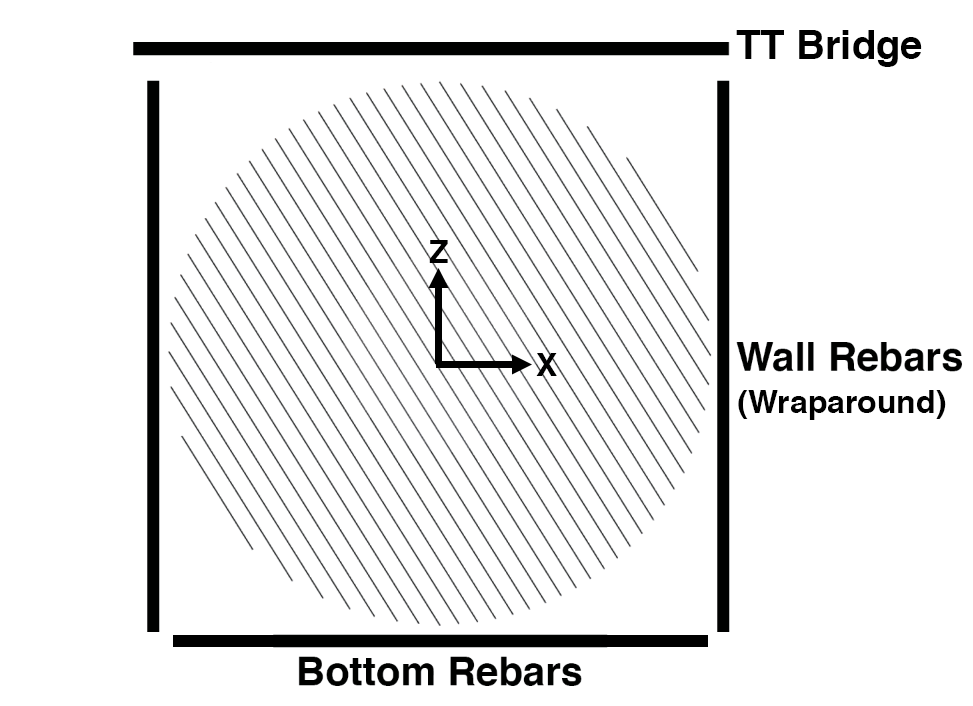}
\caption{\label{fig:locationdiagram} Side view of the  compensation coils, with the locations of the evaluated rebars and TT bridge.}
\end{figure}

A large number of carbon steel rebars is used in the wall and bottom of the water pool of the JUNO detector, and a TT bridge consisting mainly carbon steel is installed above the water pool.  In this section, we briefly describe the layout of the base and wall rebars as well as the TT bridge. For convenience of the description and simulation, we set the frame as follows: the origin sits at the center of the acrylic sphere, the $z$-axis is in the up direction, and the geomagnetic field is perpendicular to the $y$ axis. Figure \ref{fig:locationdiagram} shows the side view of the  compensation coils, the locations of the rebars and TT bridge. 

\begin{figure}
	\centering
	\includegraphics[width=\textwidth]{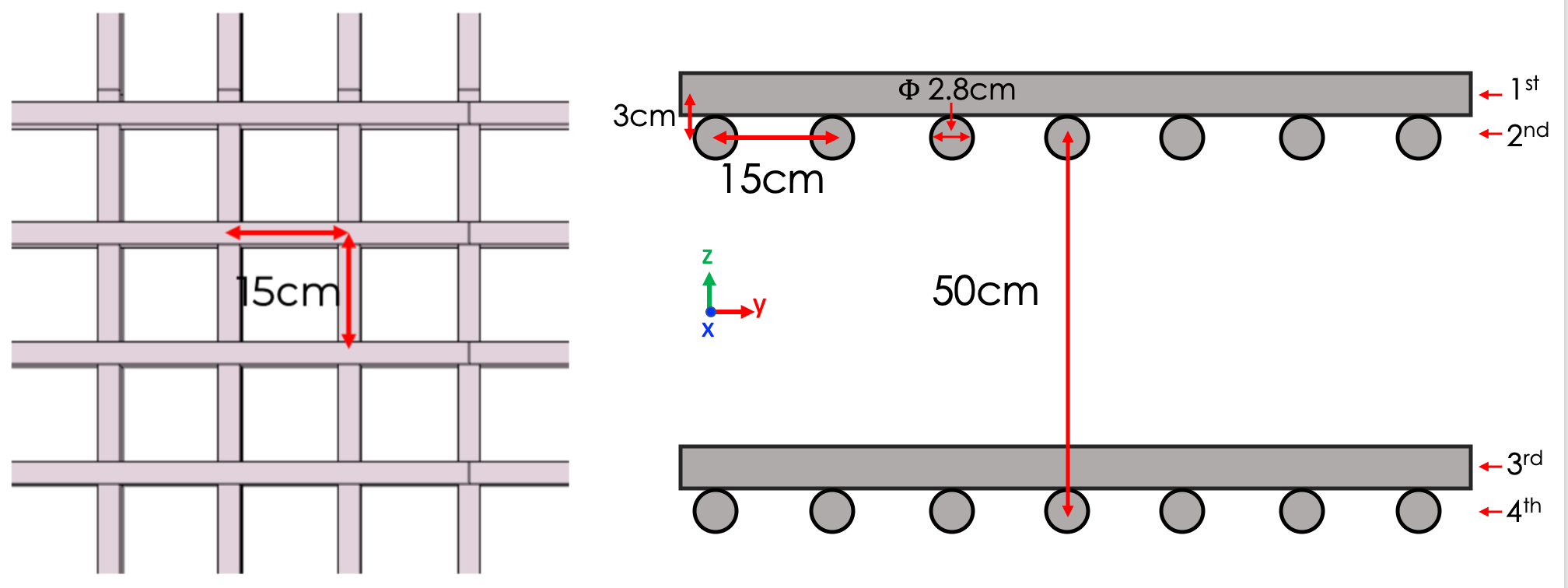}
	\caption{\label{fig:brp} (Left) Top-down view of the bottom rebars, showing the mesh structure formed by layers of rebars. (Right) Side view showing the dimensions for the 4 layers of the bottom rebars.}
	\label{distributions}
\end{figure}

The base of the water pool consists of 4 layers of rebars, where the $1^{st}$, $2^{nd}$, $3^{rd}$ and $4^{th}$ layers are located at 21.78 m, 21.81 m, 22.28 m and 22.31 m below the center of the detector (the center of the set of coils), respectively. The rebars are weaved into two layers of mesh roughly 50 cm apart in the $z$-direction, with 15 cm rebars spacings on each layer, as shown in the left panel of figure~\ref{fig:brp}. The rebars are arranged to cover the entire base of the water pool, but denser in the central part with a diameter of 27 m.

\begin{figure}[h!]
\centering 
\includegraphics[width=13cm]{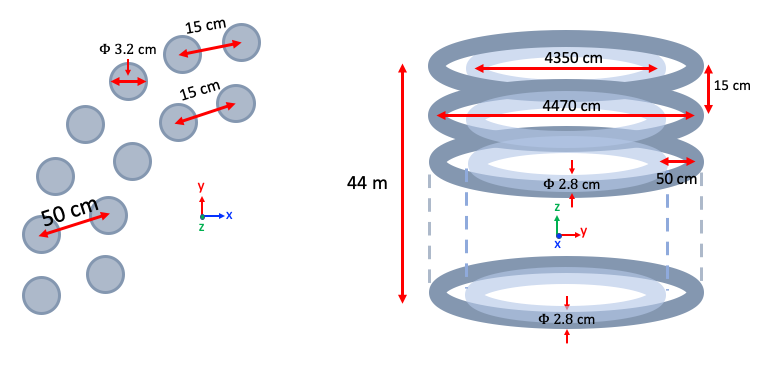}
\caption{\label{fig:rw} (Left) Dimension of the vertical rebars of the two layers.  (Right) Dimension for the horizontal rebars of the inner and outer shells.}
\end{figure}

The wall of the cylindrical water pool is lined with two layers of cylindrical meshes of 43.7 m and 44.7 m diameter extending from the bottom to the top of the concrete wall to form a bird cage. Each layer of the mesh is formed from vertical rebars with diameter of 32 mm and horizontal rebars with diameter of 28 mm, shown in Figure~\ref{fig:rw}. The vertical and horizontal rebars on each mesh are arranged 15 cm apart, same as the bottom rebars. 

\begin{figure}[htbp]
\centering 
\includegraphics[width=0.8\linewidth]{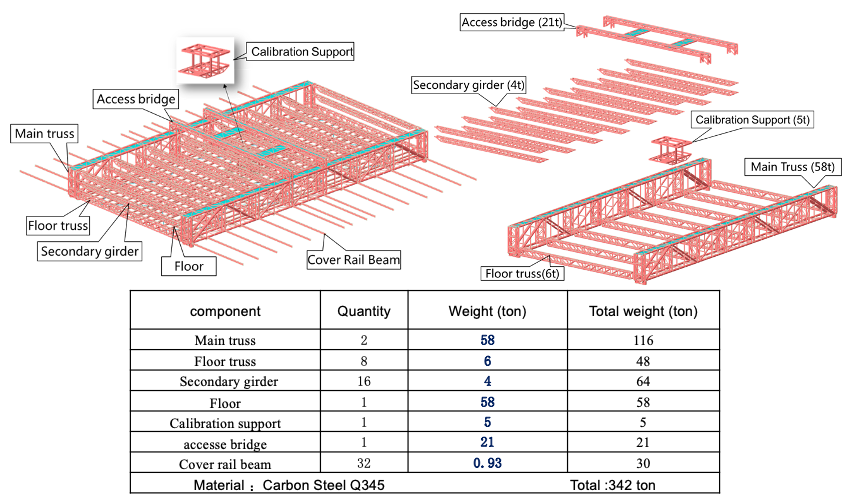}
\caption{\label{fig:wtt} The JUNO's main support structure for the TT-bridge and its weight.}
\end{figure}

As shown in Figure~\ref{fig:wtt}, the structure of the TT-bridge consists of the main truss, floor truss, secondary girder, floor, calibration support, access bridge and cover rail beam. The central components compose of 48 tons of the floor trusses, 64 tons of the secondary girders, 58 tons of the floor, 5 tons of the calibration support, 21 tons of the access bridge, and 30 tons of the cover rail beams. The two main trusses, each 58 tons, are the wing components of the main support structure. 

\section{Effect of Rebars and TT Bridge Steel}

\begin{figure}[h]
\centering 
\includegraphics[width=0.9\linewidth]{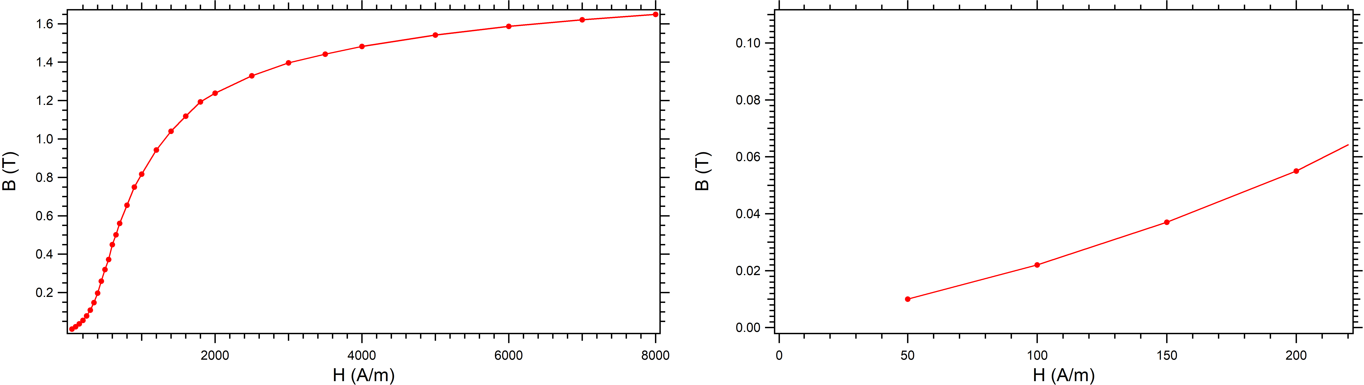}
\caption{\label{fig:bhc} B-H curve of HRB 400 carbon steel rebar across the range measured by IHEP (Left), B-H curve at low magnetic field strength (Right). The measurements have an uncertainty of $\pm1.0\%$ at 2$\sigma$} (given by NIM).
\end{figure}

We evaluate in this section the magnetic field contribution from the rebars in the base of the water pool, the rebars in the wall of the water pool, and the steel of the TT bridge in conjunction with the compensation coils. The steel used inside the concrete layer of the water pool and the TT bridge is standard structural steel of type HRB400. In our calculation, we employ the B-H curve of the HRB400 steel from the report, No. GJcc2018-0967, of The National Institute of Metrology, China (NIM)~\cite{nimWelcomeNational} as the material property for the rebars and the TT bridge. The B-H curve is as shown in Figure~\ref{fig:bhc}.

All the calculations of this work were performed with the Radia 3D magnetostatics code, developed by the Insertion Devices laboratory of the European Synchrotron Radiation Facility (ESRF) \cite{Chubar:hi3178}. The computation method used by Radia differs from commercial simulation software by utilizing the integral boundary method instead of finite element method (FEM), with extensive use of analytical expressions for the magnetic field and field integral calculations. The main advantage of Radia is its use in geometries that open to infinity, as the empty spaces do not need to be meshed, resulting in lower memory usage and CPU time compared to FEM \cite{753258}. The code is interfaced to Mathematica \cite{Mathematica}, with the pre- and post-processing also in Mathematica.

\begin{figure}[htbp]
\centering 
\includegraphics[width=0.8\linewidth]{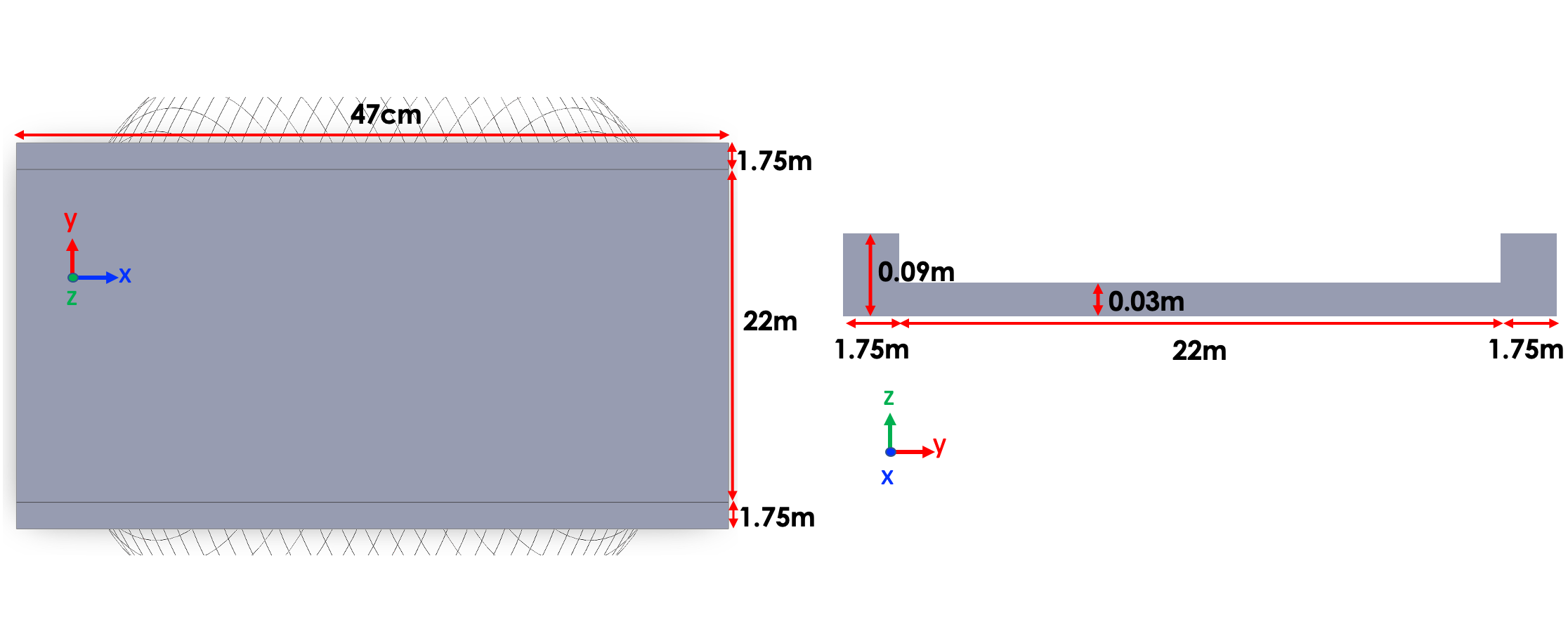}
\caption{\label{fig:dimtt} The dimension of the main support structure of the TT bridge for simulation. Geometry exaggerated for visualization purpose}
\end{figure}

The location of the TT bridge structure as shown in Figure~\ref{fig:wtt} is 1.86 m above the top of the compensation coils. Due to the complexity of the structure, the TT bridge is simplified for the calculations by compressing the geometry into three rectangular prism, maintaining the weight of the structure, resulting in the geometry shown in Figure~\ref{fig:dimtt}. By maintaining the weights of the structure, the dimension for each main truss becomes $1.75~m \times 47~m \times 0.09~m$ while the central components become $22~m \times 47~m \times 0.03~m$.

The wall rebars are implemented by simulating the horizontal and vertical rebars that wrap around to form a bird cage of 45 m in diameter and 44 m in height, as shown in Figure \ref{fig:locationdiagram}. However, the numbers of rebars are reduced to $1/4$ of the actual numbers, by increasing the spacing between rebars from 15 cm to 30 cm in both horizontal and vertical direction. The reduction in number of rebars is compensated by doubling the volume of each individual rebars. The bottom rebars are implemented by simulating 
the 45 m water pool bottom, with roughly 1,500 full-size rebars spanning the entire length of the bottom. All the rebars are segmented into smaller pieces for accurate magnetic field simulation.

The calculations take into account the geomagnetic field at the location of JUNO site, the magnetic field generated by the compensation coils and the magnetization of the rebars. Note that the main axis of the geomagnetic field compensation coils goes through the origin (center of the detector) and lays on the $xz$ plane.

\begin{figure}[h]
\centering 
\includegraphics[width=0.75\linewidth]{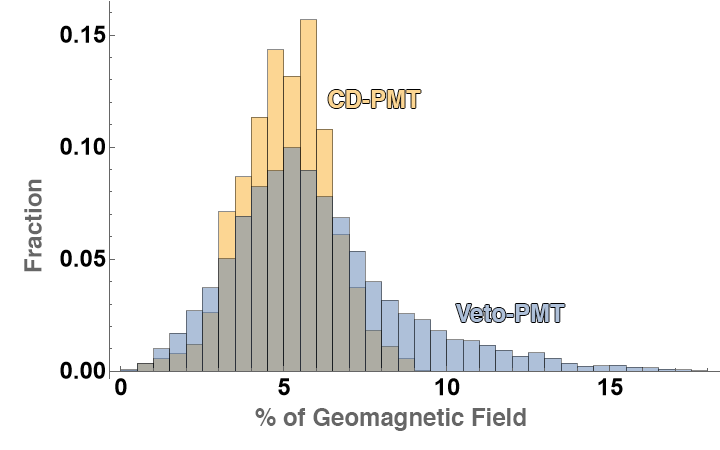}
\caption{\label{fig:histcombined} Comparison of the residual magnetic field strength at the locations of CD and Veto PMTs. }
\end{figure}

The resulting fields are sampled with 20,000 points uniformly spaced on each of the spherical surfaces that represent the CD-PMTs and Veto-PMTs and plotted below. Figure \ref{fig:histrebars} shows the effect that rebars and TT bridge have on the residual magnetic field strength. The yellow histograms are for the system with only the compensation coils and the geomagnetic field at JUNO's site, while the blue histograms are the results when all the rebars and TT bridge are added. It is found that without the influence of construction steel, the residual field would be below $3\%$ of the geomagnetic field, but the addition of steel greatly increases the residual field. As shown in  Figure \ref{fig:histcombined}, most of the CD-PMTs experience residual fields around $5\%$ of the geomagnetic field strength, and all experiences below the acceptable limit of $10\%$. The Veto-PMTs predominantly experience fields below $10\%$ of the geomagnetic field strength, with a small number of PMTs experiencing higher fields upto $18\%$.
\begin{figure}[htbp]
\centering 
\includegraphics[width=1\linewidth]{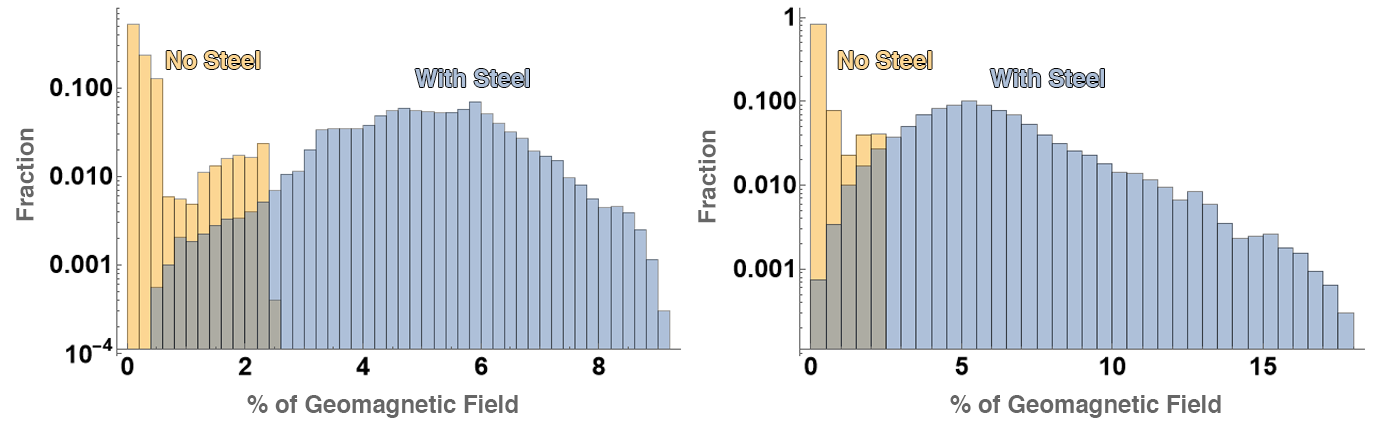}
\caption{\label{fig:histrebars} Comparison of the residual magnetic field strength at the locations of CD-PMTs (left), and Veto-PMTs (right), between systems without any steel (only compensation coils and geomagnetic field), and with rebars and TT bridge. The vertical axis is plotted in log-scale for clarity.}
\end{figure}

To visualize the magnetic field strength at the locations of the CD-PMTs and Veto-PMTs, the 3-dimensional surface on the central detector sphere is projected onto 2-dimensional plots. The coordinates of the following plots are defined relative to the central detector sphere, with the center of the sphere at (0, 0, 0). 

\begin{figure}[htbp]
\centering 
\includegraphics[width=1\linewidth]{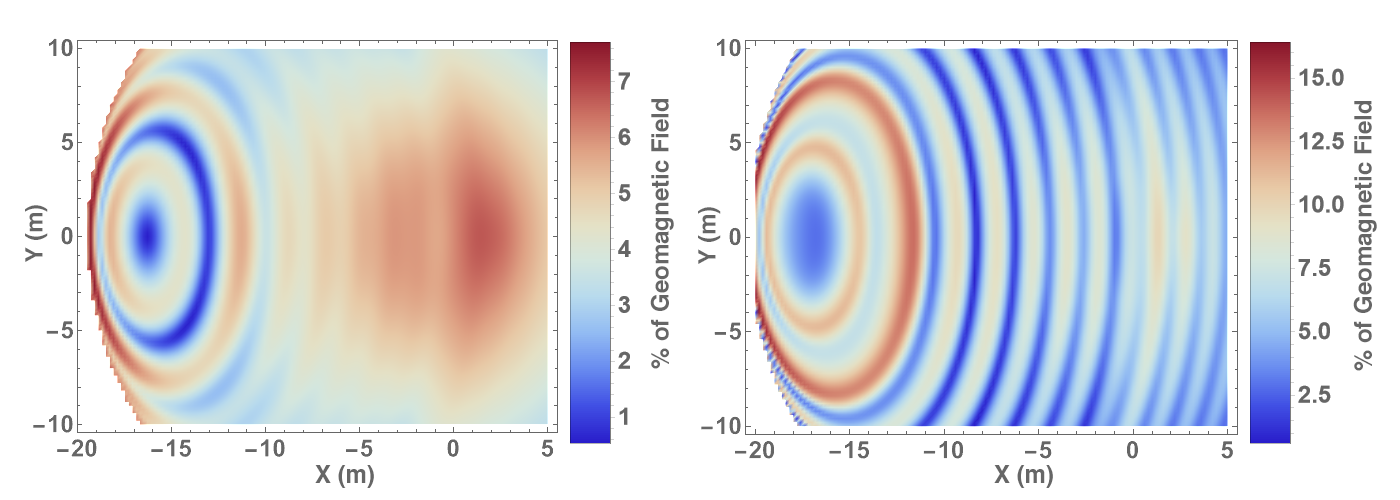}
\caption{\label{fig:botcombined} Plots show the residual magnetic field strength relative to the geomagnetic field strength in the CD-PMT region (left) and in the Veto-PMT region (right). The point (0, 0) is the lowest point of the central detector sphere, closest to the bottom of the water pool.
The center of the solid blue circles corresponds to the main axis of the coils. }
\end{figure}
Our simulations reveal that only the PMTs close to the bottom of the water pool, right below the TT bridge, and around the main axis of the coils experience considerable magnetic fields. Figure~\ref{fig:botcombined} shows the residual magnetic field strength in the region near the bottom of the water pool. The coordinate $(0,0)$ of the plots represents the lowest point of the central detector. It is found in Figure~\ref{fig:botcombined} that for the CD-PMT region the maximum residual magnetic field strength is about 7.6\% of the geomagnetic field, and in the Veto-PMT region the maximum is 16\%. The simulation results reveal that for the Vedo-PMTs, it is not the ones closest to the bottom of the pool but the ones close to the main axis of the coils experience the maximum magnetic field since the combined magnetic field around the main axis of the coils is rather strong.  

\begin{figure}[htbp]
\centering 
\includegraphics[width=1\linewidth]{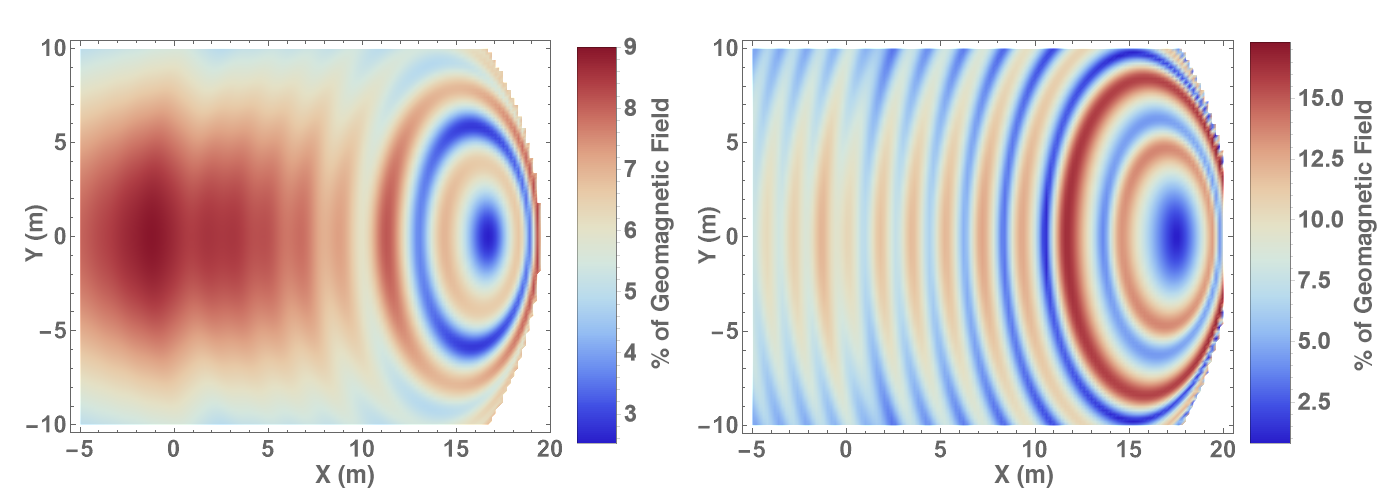}
\caption{\label{fig:ttcombined}The residual magnetic field strength relative to the geomagnetic field strength in the CD-PMT region (left) and in the Veto-PMT region (right). The point (0, 0) is the highest point of the central detector sphere, closest to the TT bridge. The center of the solid blue circles corresponds to the main axis of the coils.  } 
\end{figure}

Figure~\ref{fig:ttcombined} shows the residual magnetic field strength in the region under the TT bridge. The coordinate $(0,0)$ of the plots represents the highest point of the central detector. We see that the residual field for the Veto-PMT region is slightly higher than the bottom region shown previously, with the maximum value reaching up to $17\%$. The result on the CD-PMT region, as shown in the left pane, is in the same track, with the maximum residual intensity at $9 \%$ of the geomagnetic field.
\begin{figure}[htbp]
\centering 
\includegraphics[width=1\linewidth]{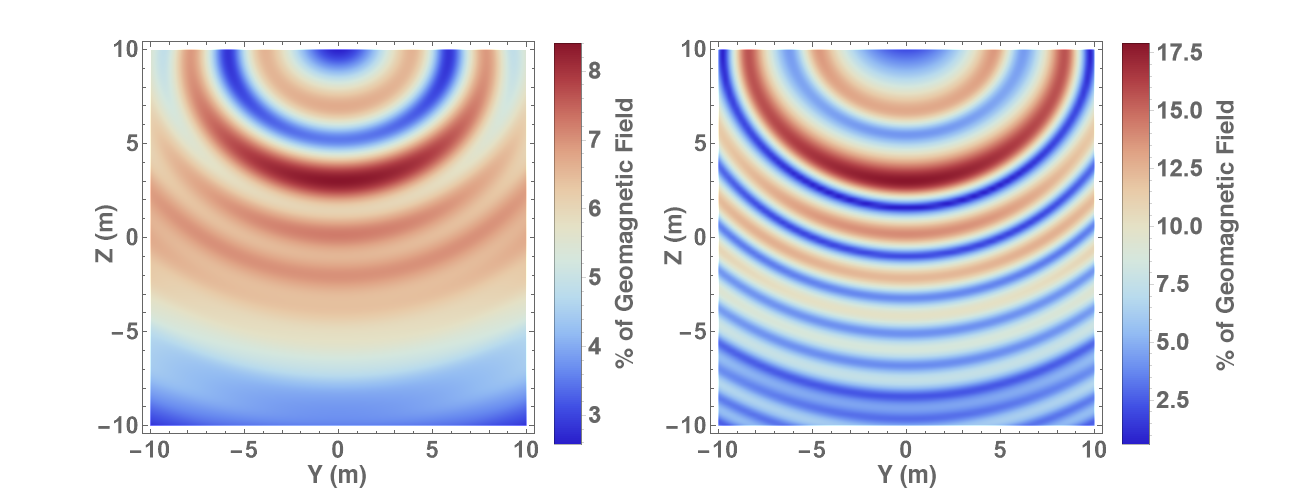}
\caption{\label{fig:wallcombined} The residual magnetic field strength relative to the geomagnetic field strength in the CD-PMT region (left) and in the Veto-PMT region (right). The point (0, 0) is the widest part of the central detector sphere, closest to the water pool wall. The center of the solid blue circles corresponds to the main axis of the coils. }
\end{figure}

Figure~\ref{fig:wallcombined} shows the residual magnetic field in the region near the main axis of the compensation coils, where the rebars are subject to a strong combined magnetic field intensity from the compensation coils and geomagnetic field. The coordinate $(0,0)$ of the plots represents the side-most point of the central detector on the $xz$-plane. The maximum of the residual magnetic field is almost $9\%$ for the CD-PMT region, and $18\%$ for the Veto-PMT region. The calculation also reveals that in the region where the largest compensation coils are accommodated, the residual magnetic field is minimally affected by the addition of the wall rebars.

\section{Conclusion}

In summary, our simulations demonstrate that despite the presence of carbon steel structures such as rebars and the TT bridge within the central detector's vicinity, the residual magnetic field experienced by the PMTs remains within the acceptable limit established by the JUNO experiment of 10\% for CD-PMTs and 20\% for Veto-PMTs, compared to the geomagnetic field. The maximum magnetic fields experienced by the CD-PMTs are 7.6\% above the bottom rebars, 9\% below the TT bridge, and 9\% by the wall. The Veto-PMTs experience maximums of 16\% above the bottom rebars, 17\% below the TT bridge, and 18\% by the wall. These findings indicate that the residual magnetic field will have minimal impact on the PMTs' photon detection efficiency, based on the real-world measurements in Figure~\ref{fig:pmteff}. 
\\

\section*{Acknowledgements}
This research has received funding support from the NSRF via the Program Management Unit for Human Resources \& Institutional Development, Research and Innovation [grant number B37G660014]. JS and YY acknowledge support from Suranaree University of Technology (SUT) - Full-time Master Researcher (contract no. Full-time 61/05/2562). This work was supported by the Chinese Academy of Sciences,
the National Key R\&D Program of China and the Key Intergovernmental Specialities in International Scientic and Technological Innovation Cooperation, MOST (No. 2017YFE0132500).




\bibliography{SciPost_Example_BiBTeX_File.bib}

\end{document}